\documentclass[twocolumn]{aastex631}
\received{September 1, 2023}

\submitjournal{ApJ}

\begin{document}

\defcitealias{Sand+04}{S04} 

\newcommand{\gtot}{\gamma_{{\rm tot}}}
\newcommand{\gdm}{\gamma_{{\rm DM}}}
\newcommand{\rtwo}{r_{200}}
\newcommand{\rs}{r_{\rho}}
\newcommand{\rv}{r_{\Delta}}
\newcommand{\rfive}{r_{500}}
\newcommand{\rone}{r_{100}}
\newcommand{\rnu}{r_{\nu}}
\newcommand{\cnu}{c_{\nu}}
\newcommand{\rdue}{r_2}
\newcommand{\mv}{M_{\Delta}}
\newcommand{\mtwo}{M_{200}}
\newcommand{\mfive}{M_{500}}
\newcommand{\prmt}{M_{p}(0.2)}
\newcommand{\prmf}{M_{p}(0.5)}
\newcommand{\mpr}{M_p(<R)}
\newcommand{\ks}{\rm{km} \, \rm{s}^{-1}}
\newcommand{\msun}{M_{\odot}}
\newcommand{\lsun}{L_{\odot}}
\newcommand{\ctwo}{c_{200}}
\newcommand{\slos}{\sigma_{\rm{los}}}
\newcommand{\ab}[1]{\textcolor{blue}{\bf #1}}
\newcommand{\SE}[1]{\textcolor{orange}{\bf #1}}
\newcommand{\rr}[1]{{\bf \textcolor{red}{\bf #1}}}

\title{CLASH-VLT: The inner slope of the MACS J1206.2$-$0847 dark matter density profile}

\correspondingauthor{Andrea Biviano}
\email{andrea.biviano@inaf.it}

\author[0000-0002-0857-0732]{Andrea Biviano}
\affiliation{INAF-Osservatorio Astronomico di Trieste, via G.B. Tiepolo 11, 34143 Trieste, Italy}
\affiliation{IFPU-Institute for Fundamental Physics of the Universe,
via Beirut 2,
34014 Trieste, Italy}

\author[0000-0001-5654-7580]{Lorenzo Pizzuti}
\affiliation{CEICO, Institute of Physics of the Czech Academy of Sciences, Na Slovance 2, 182 21 Praha 8, Czechia}
\affiliation{Dipartimento di Fisica G. Occhialini, Universit\'a degli Studi di Milano Bicocca, Piazza della Scienza 3, I-20126 Milano, Italy}

\author[0000-0001-9261-7849]{Amata Mercurio}
\affiliation{ Department of Physics of the University of Salerno - Via Giovanni Paolo II, 132, 84084, Fisciano (SA), Italy \label{unisa}}
\affiliation{INAF-Astronomical Observatory of Capodimonte, Salita Moiariello 16, I-80131 Napoli, Italy\label{oana}}

\author[0000-0003-1337-5269]{Barbara Sartoris}
\affiliation{Universit\"ats-Sternwarte M\"unchen, Fakult\"at f\"ur Physik, Ludwig-Maximilians Universit\"at, Scheinerstr. 1, 81679 M\"unchen, Germany}
\affiliation{INAF-Osservatorio Astronomico di Trieste, via G.B. Tiepolo 11, 34143 Trieste, Italy}

\author[0000-0002-6813-0632]{Piero Rosati}
\affiliation{Dipartimento di Fisica e Scienze della Terra, Università degli Studi di Ferrara, Via Saragat 1, I-44122 Ferrara, Italy}
\affiliation{INAF -- OAS, Osservatorio di Astrofisica e Scienza dello Spazio di Bologna, via Gobetti 93/3, I-40129 Bologna, Italy}

\author[0000-0003-4117-8617]{Stefano Ettori}
\affiliation{INAF -- OAS, Osservatorio di Astrofisica e Scienza dello Spazio di Bologna, via Gobetti 93/3, I-40129 Bologna, Italy}
\affiliation{INFN, Sezione di Bologna, viale Berti Pichat 6/2, 40127 Bologna, Italy}

\author[0000-0003-1861-1865]{Marisa Girardi}
\affiliation{Dipartimento di Fisica, Universit\'a di Trieste, via Tiepolo 11, 34143 Trieste, Italy}

\author[0000-0002-5926-7143]{Claudio Grillo}
\affiliation{Dipartimento di Fisica, Università degli Studi di Milano, Via Celoria 16, I-20133 Milano, Italy}
\affiliation{INAF - IASF Milano, via A. Corti 12, I-20133 Milano, Italy}

\author[0000-0001-6052-3274]{Gabriel B. Caminha}
\affiliation{Technical University of Munich, School of Natural Sciences, Department of Physics, James-Franck-Stra{\ss}e 1, 85748 Garching, Germany}


\author[0000-0001-6342-9662]{Mario Nonino}
\affiliation{INAF-Osservatorio Astronomico di Trieste,
via G.B. Tiepolo 11, 34143 Trieste, Italy}



\begin{abstract}

The inner slope ($\gdm$) of the dark matter (DM) density profile of cosmological halos carries information about the properties of DM and/or baryonic processes affecting the halo gravitational potential. Cold DM cosmological simulations predict steep inner slopes, $\gdm \simeq 1$. We test this prediction on the MACS J1206.2$-$0847 cluster at redshift $z=0.44$, whose DM density profile was claimed to be cored at the center. We determine the cluster DM density profile from 2 kpc from the cluster center to the virial radius ($\sim 2$ Mpc), using the velocity distribution of $\simeq 500$ cluster galaxies and the internal velocity dispersion profile of the Brightest Cluster Galaxy (BCG), obtained from \texttt{VIMOS@VLT} and \texttt{MUSE@VLT} data. We solve the Jeans equation of dynamical equilibrium using an upgraded version of the MAMPOSSt method. The total mass profile is modeled as a sum of a generalized-NFW profile that describes the DM component, allowing for a free inner slope of the density profile,  a Jaffe profile that describes the BCG stellar mass component, and a non-parametric baryonic profile that describes the sum of the remaining galaxy stellar mass and of the hot intra-cluster gas mass. Our total mass profile is in remarkable agreement with independent determinations based on X-ray observations and strong lensing. We find $\gdm=0.7_{-0.1}^{+0.2}$ (68\% confidence levels), consistent with predictions from recent $\Lambda$CDM cosmological numerical simulations. 

\end{abstract}



\section{Introduction} \label{s:intro}
Investigating the mass distribution of cosmological halos is important
to understand the halo assembly process through the interplay of Dark
Matter (DM) and baryons \citep[e.g.,][]{Blumenthal+86,EZSH01,LO10,Correa+15-I}, and
to constrain the properties of DM itself and of gravitational
interactions \citep[e.g.,][]{ABG02,Markevitch+04,Sartoris+14,PSUS22}. Cold DM-only
cosmological simulations find that DM halo mass density profiles follow the universal NFW model,
  \begin{equation}
  \rho(r) \propto (r/\rs)^{-1} \, (1+r/\rs)^{-2},\label{e:nfw}
  \end{equation}
over a wide range of halo masses, from the center to
the virial radius \citep{NFW96,NFW97}. The
logarithmic slope of the NFW model changes from $\gdm = -d \log \rho / d \log r
= 1$ at $r=0$, to $3$ at large radii, and its
characteristic radius, $\rs$, corresponds to the radius $\rdue$ where $\gdm=2$.

In following studies, based on simulations with higher resolution than the original one by \citet{NFW96}, the universality of halo $\rho(r)$ has been questioned \citep[e.g.,][]{RPV07,DelPopolo10}, and a generalized form of the initial NFW model (gNFW hereafter) has been proposed by \citet{WTS01},
\begin{equation}
  \rho \propto (r/\rs)^{-\gdm} \, (1+r/\rs)^{-3+\gdm}. \label{e:gnfw}
\end{equation}
In this case, $\rs \equiv \rdue/(2-\gdm)$.
The gNFW model allows a free value of the inner slope of the halo DM profile, $\gdm$. Based on the recent \texttt{C-EAGLE} hydrodynamic simulations \citep{Barnes+17}, \citet{He+20} constrain the average inner slope of the DM density profile of cluster-size halos, and its scatter, finding $\gdm=1.07 \pm 0.06$.

The inner slope $\gdm$ of halo $\rho(r)$ can depend on the nature of DM. Numerical simulations have shown that there are several alternative models to the Cold DM (CDM) scenario, such as warm, fuzzy, decaying, and self-interacting DM, that can produce halos with $\gdm<1$ \citep{BOT01,HBG00,PMK10,SS00}. Self-interacting DM halo profiles, in particolar, are well fit by cored isothermal profiles \citep{RME21}. 

The inner structure of halos and the value of $\gdm$ can also differ from the predictions of CDM-only simulations because of collisional processes that concern the baryonic components. In particular, $\gdm>1$ can result from the process of adiabatic contraction caused by the central condensation of cooled gas, and from the process of mass accretion \citep{Blumenthal+86,Gnedin+04,Laporte+12,DK14,Schaller+15}. On the other hand, the processes of dynamical friction and AGN feedback can flatten the central slope of the DM profile \citep[$\gdm<1$,][]{EZSH01,ElZant+04,MTMW12,RFGA12,Peirani+17}. 

Clusters of galaxies are particularly well suited for the study of the DM density profiles, since their mass budget is dominated by DM at most radii \citep[see, e.g.][]{BS06}, at variance with galaxies whose mass profile at the center is dominated by baryons \citep{RME21}. The cluster
mass profile can be constrained through X-ray and Sunyaev-Zel'dovich \citep{SZ69} observations of the hot intra-cluster medium (ICM hereafter), weak gravitational lensing, and the kinematics of cluster galaxies \citep[e.g.,][]{Pratt+19}. The best probes of the cluster gravitational potential close to its center are strong gravitational lensing \citep[e.g.][]{MFK93,Zitrin+12} and the internal kinematics of the brightest cluster galaxy \citep[BCG, see e.g][]{Dressler79,Kelson+02}. The output of all these measurements is the cluster total mass profile, to which one must subtract the contribution of baryons to extract the DM profile. Most baryons outside the very cluster center are contributed by the ICM, while the BCG stellar mass dominates the baryon budget close to the center, where it makes a non-negligible, and in some cases dominant, contribution to the total mass budget \citep[see, e.g.,][]{BS06}. Therefore, measuring the BCG stellar mass profile is fundamental to obtain a reliable estimate of $\gdm$.

Previous observational determinations of $\gdm$ for clusters of galaxies span a wide range of values. \citet{Kelson+02} studied the dynamics of the cluster A2199 by combining the kinematics of the BCG and intra-cluster light stars with that of the cluster members, and assuming isotropic velocity distributions. They found that a cored DM halo reproduces the observed kinematics better than a NFW profile. Based on a strong lensing analysis of three clusters, \citet{LBJ22} find that cored inner mass density profiles are favored over cuspy models. Using a combination of strong gravitational lensing and BCG kinematics, \citet{Newman+13a} estimate $\gdm=0.50 \pm 0.13$ averaging over seven clusters, in agreement with previous results by the same collaboration \citep{STE02,Sand+08,Newman+09,NTES11}. All seven clusters, and another two from a previous investigation \citep[][hereafter S04]{Sand+04}, have $\gdm<1$ with various levels of statistical confidence.  A larger value is found by \citet{Annunziatella+17}, who estimate $\gdm=1.36 \pm 0.01$ for the cluster MACS~J0416$-$2403, based on BCG kinematics, X-ray and strong lensing. However, their estimate is based on the assumption of a single power-law mass profile for de-projection, and this assumption could lead to an over-estimate of $\gdm$ if the intrinsic 3D mass profile steepens with radius, as expected for clusters of galaxies. By combining the BCG with the cluster kinematics (as traced by its member galaxies), \citet{Sartoris+20}, find $\gdm=0.99 \pm 0.04$ for the cluster Abell~S1063. 

The mass profile inner slope of the cluster MACS J1206.2$-$0847 (MACS~1206 hereafter), has been determined by five different studies. \citet{Umetsu+12}, \citet{Caminha+17}, and \citet{Young+15} have determined the inner slope of the {\it total} mass density profile, $\gamma_{{\rm tot}}$, the first two based on strong lensing, and the latter based on the Sunyaev-Zel'dovich emission. The strong lensing analyses found $\gamma_{{\rm tot}} \approx 0.9$-1.0, while the analysis of \citet{Young+15} found a smaller value, 0.7 (no error bars provided). \citetalias{Sand+04} and \citet{MGDHL20} have determined $\gdm$ based on strong lensing, in combination with the internal kinematics of the BCG in the case of \citetalias{Sand+04}. Of the two values, one is given without an error estimate \citep{MGDHL20} and the other is zero, and significantly smaller than the NFW value \citepalias{Sand+04}. 
The result of \citetalias{Sand+04} is supported by the strong lensing analysis of \citet{LBJ22}, who find that a cored inner mass density profile is a better fit to the data than a cuspy model.

Given the rather extreme $\gdm$ value measured for MACS~1206 by \citetalias{Sand+04}, and supported by the analysis of \citet{LBJ22},
it is interesting to have a new, independent determination of it with a kinematic data set of superior quality. In this paper, we apply the procedure of \citet{Sartoris+20} to
our new data for the cluster galaxy redshifts and the BCG velocity dispersion profile, that come from the CLASH-VLT ESO Large Programme \citep[D 186.A-0798, P.I. P. Rosati,][]{Rosati+14} and from additional archival observations obtained with the integral field spectrograph \texttt{MUSE@VLT} \citep{Caminha+17}.

The structure of this paper is the following. In Sect.~\ref{s:data} and Sect.~\ref{s:method} we describe our data set, and the method of analysis, respectively. We provide our results in Sect.~\ref{s:res} with the relevant discussion in Sect.~\ref{s:disc}. In Sect.~\ref{s:summ} we give a summary of our results and our conclusions. Throughout this paper we adopt the following cosmological parameters: $\Omega_m=0.3, \Omega_{\Lambda}=0.7, H_0=70$ km~s$^{-1}$~Mpc$^{-1}$. At the cluster redshift, $z=0.44$, 1 arcmin corresponds to 340 kpc.

\section{The data set} \label{s:data}
Our data set consists of 3110 sources with measured redshift, 2650 obtained with \texttt{VIMOS@VLT} within the CLASH-VLT ESO Large Programme \citep[D 186.A-0798, P.I. P. Rosati,][]{Rosati+14}, 410 obtained with \texttt{MUSE@VLT} \citep{Caminha+17}\footnote{Based on the GTO programs 095.A-0181(A), 097.A-0269(A) (P.I. J. Richard)}, 14 observed with \texttt{FORS@VLT} \citep{Presotto+14}, 11 observed with \texttt{IMACS-GISMO} at the Magellan telescope (Daniel Kelson, priv. comm.) and another 25 gathered from the literature \citep{Lamareille+06, Jones+04,Ebeling+09}. Details on the construction of the spectroscopic catalogue can be found in \citet{Balestra+16}. The uncertainties in the redshift measurements correspond to uncertainties in the rest-frame velocities of  cluster galaxies of 153, 75, 15 $\ks$ for the measurements obtained with \texttt{VIMOS-LR, VIMOS-MR,} and \texttt{MUSE}, respectively. In addition, we have obtained very accurate \texttt{MUSE} measurements of the surface brightness and velocity dispersion profiles of the BCG.  

\subsection{The selection of cluster members} \label{ss:members}
\begin{figure}[ht]
\includegraphics[width=0.45\textwidth]{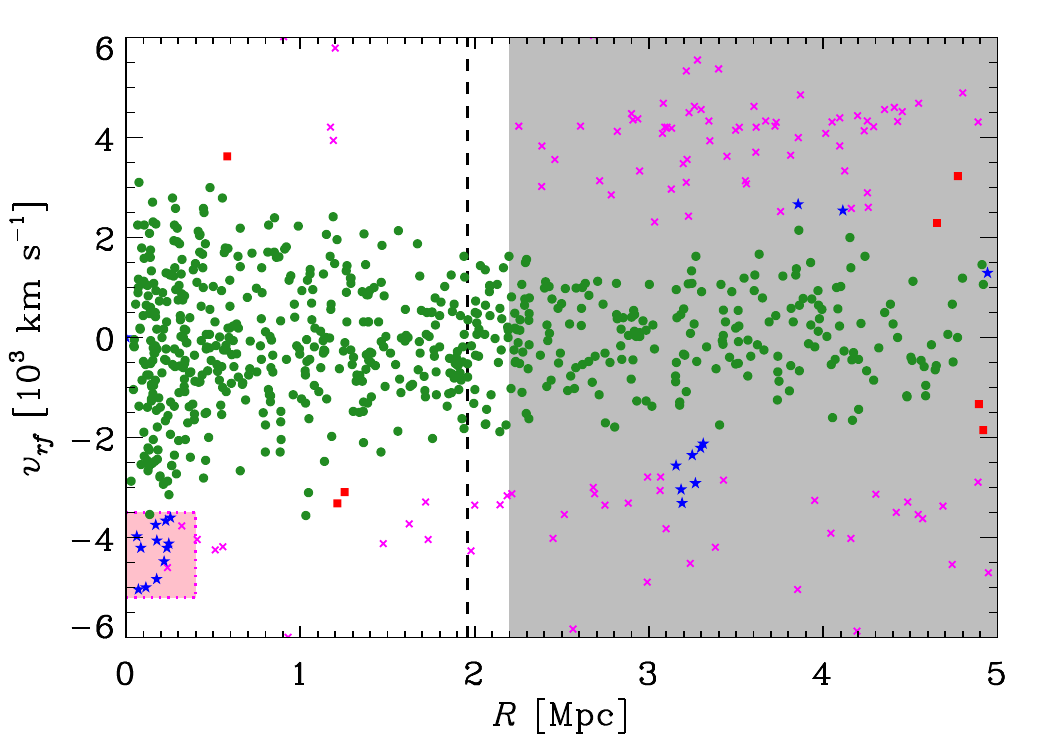}
\caption{Rest-frame line-of-sight velocity ($\ks$) vs. projected cluster-centric distance (kpc) of galaxies in the cluster region. Red squares indicate members selected by the \texttt{CLUMPS} method alone (7 galaxies), blue stars indicate members selected by the \texttt{P+G} method alone (31 galaxies), green dots indicate members selected by both \texttt{P+G} and \texttt{CLUMPS} (680 galaxies), and magenta crosses indicate galaxies that are not selected as members by neither of the two methods (2392 galaxies). The black vertical dashed line indicates the value of the virial radius, 1.96 Mpc, according to \citet{Umetsu+12}. Galaxies in the grey regions are excluded from the dynamical analysis (see Sect.~\ref{s:method}). The pink region within dotted segments is the region of phase-space occupied by a likely foreground group of galaxies \citep[discussed in Sect.~\ref{s:res} and previously identified by][]{Young+15}.}
\label{f:members}
\end{figure}

\begin{figure}[ht]
\begin{center}
\includegraphics[width=0.45\textwidth]{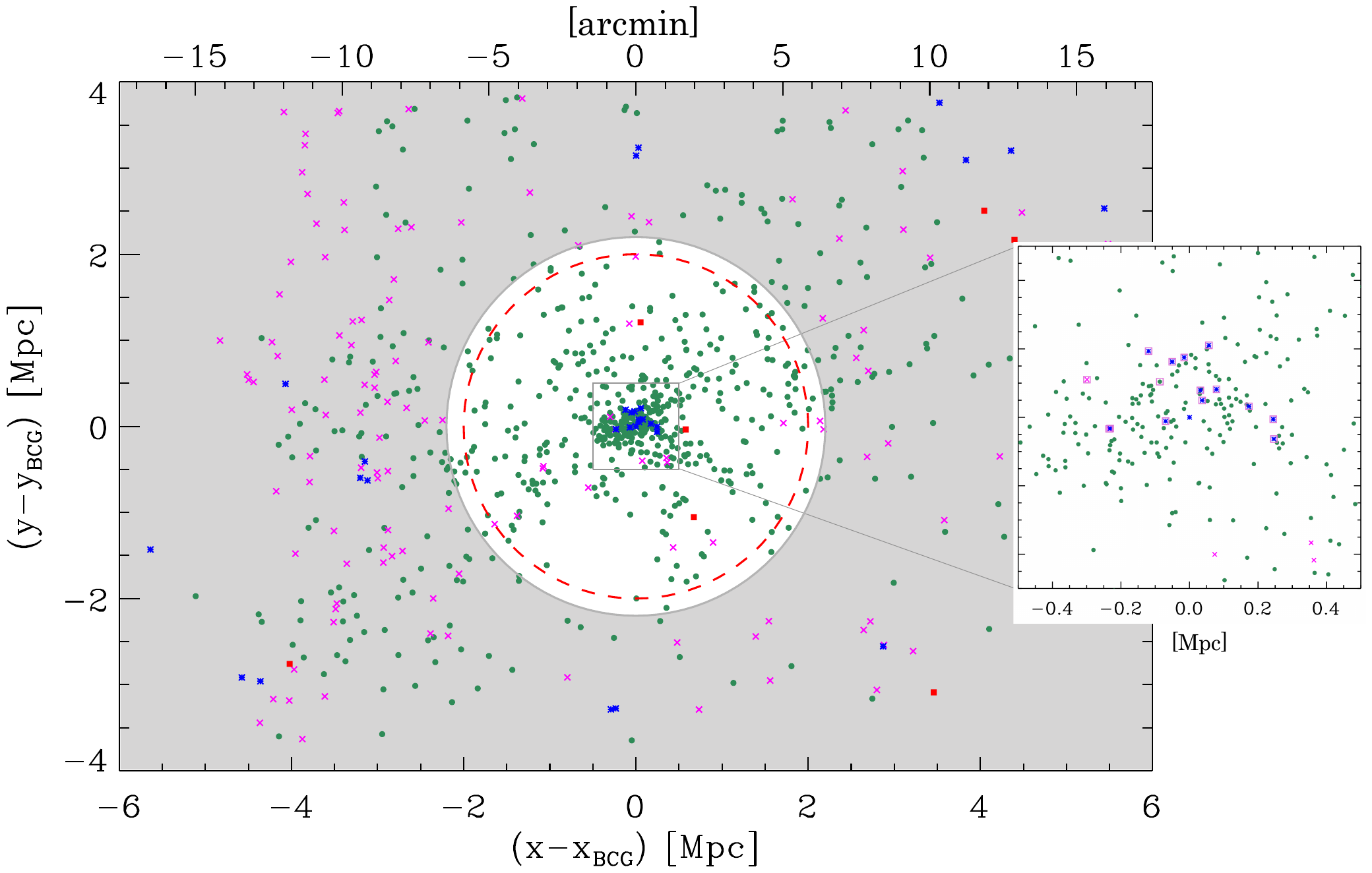}
\end{center}
\caption{Spatial distribution of member and non-member galaxies shown in the projected phase-space diagram of Fig.~\ref{f:members}, with the same symbols and colors (all other spectroscopic galaxies in the field are not shown). The dashed red circle represents the virial radius, while the white area includes galaxies used for the dynamical analysis. The inset is a blow-up of the central Mpc, where the pink open boxes mark the galaxies in the foreground group (see pink region in Fig.~\ref{f:members}).}
\label{f:map}
\end{figure}

We adopt the BCG position as the cluster center, both in spatial coordinates,
$\alpha_{\mathrm{J2000}} = 12^{\mathrm{h}}06^{\mathrm{m}}12\fs15,
\delta_{\mathrm{J2000}} = -8\degr 48\arcmin 3\farcs4$, and in redshift, $z=0.4398$. 
The BCG position is within 13 kpc of the X-ray peak position and the center of
mass determined by the gravitational lensing analysis \citep{Umetsu+12}.
We considered three independent techniques for the selection of cluster members based on their location in the cluster projected phase-space,
\begin{enumerate}
    \item \texttt{CLUMPS}, based on the identification of peaks in the velocity distribution of galaxies as a function of the cluster-centric distance \citep[see][for a full description of the method]{Biviano+21};
    \item \texttt{P+G}, based on the identification of gaps in the velocity distribution of galaxies as a function of the cluster-centric distance \citep[see][for a full description of the method]{Girardi+11,Biviano+13}.
    \item \texttt{Clean}, based on the estimate of the line-of-sight velocity dispersion profile resulting from assuming a NFW mass density profile and a given velocity anisotropy profile \citep[see][for a full description of the method]{MBB13}.
\end{enumerate}
As explained in Sect.~\ref{s:method}, we restrict our dynamical analysis to radii $R \leq 2.2$ Mpc, that is excluding galaxies (1842 in total) in the grey region in Fig.~\ref{f:members}. In this inner region, \texttt{CLUMPS}, \texttt{P+G}, and \texttt{Clean} identify 476, 485, and 482 member galaxies, respectively. In the following, we compare the \texttt{CLUMPS} and \texttt{P+G} samples, while we do not consider the \texttt{Clean} sample because it is intermediate between the two. The distribution of galaxies in the cluster projected phase-space and their spatial distribution are
shown in Fig.~\ref{f:members} and Fig.~\ref{f:map}, respectivly. In both figures we distinguish interlopers and cluster members selected by \texttt{CLUMPS}  and \texttt{P+G}.

The difference in the number of selected members between the \texttt{CLUMPS} and \texttt{P+G} methods is $<2$\%, mostly due to a few galaxies at small projected distances and relatively large negative line-of-sight velocities from the cluster center. We used the Kernel Mixture Modelling (\texttt{KMM}) algorithm \citep{McLB88,ABZ94} to check for bimodality in the velocity distributions of the two samples of members, as a possible indication of the presence of a group of galaxies in the foreground or background of the cluster. Only for the \texttt{P+G} sample we found significant evidence for bimodality. The twelve galaxies near the cluster center that \texttt{P+G} identifies as members and \texttt{CLUMPS} does not (blue stars within the pink region in Fig.~\ref{f:members}) are assigned by \texttt{KMM} a probability of 90\% to belong to a different group than the main cluster. In Sect.~\ref{s:res} we provide additional evidence that these galaxies are indeed members of a foreground group, that was previously identified by \citet{Young+15}.

Given the \texttt{KMM} results, we consider the \texttt{CLUMPS} sample of members our reference sample.
To assess the impact of a different members selection on our results, we nevertheless perform our dynamical analysis also on the \texttt{P+G} sample of members.

\subsection{The BCG} \label{ss:bcg}
\begin{figure}
\begin{center}
\includegraphics[width=0.45\textwidth]
{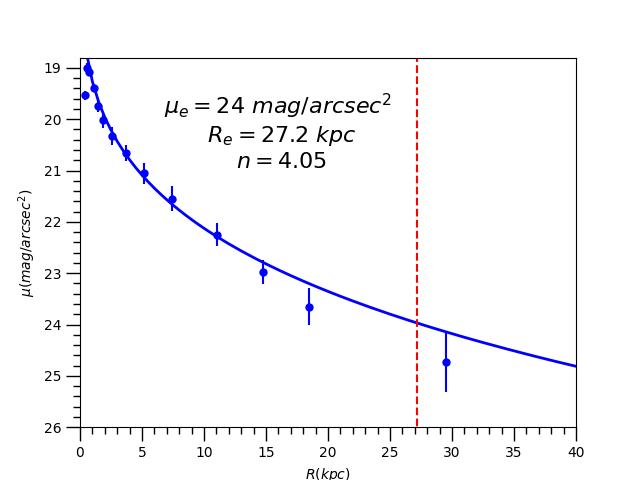}
\end{center}
\caption{Best-fit S\'ersic profile (solid blue line) to the BCG obtained by using galfit (see text). Blue points represent the surface brightnesses obtained measuring aperture magnitudes, without applying corrections for PSF to guide the eye for the fit. The vertical dashed red line shows the position of the effective radius, $R_e=27.2$ kpc.
}
\label{f:BCGsersic}
\end{figure}

\begin{figure}[ht]
\includegraphics[width=.45\textwidth]{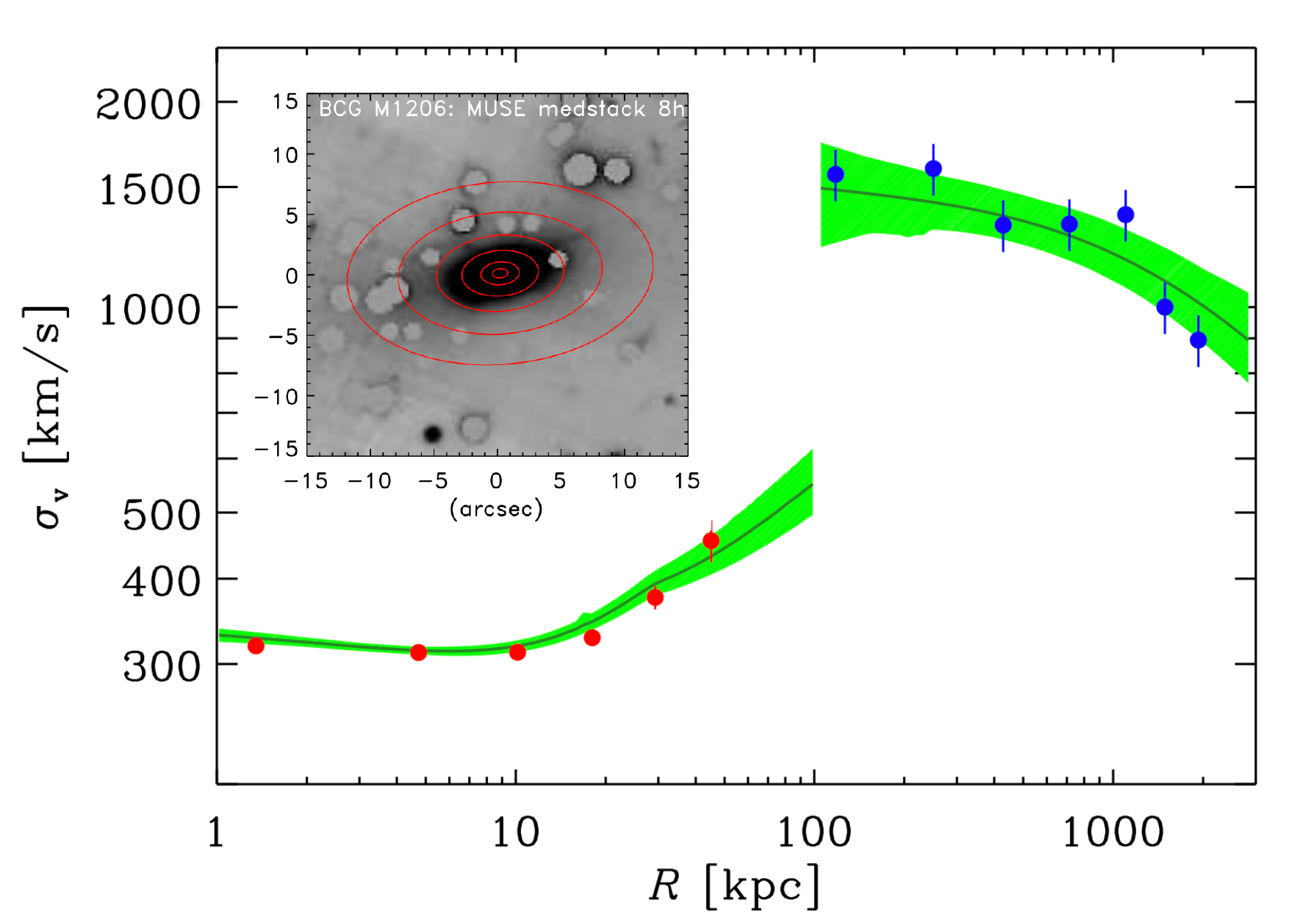}
\caption{Observed l.o.s. velocity dispersion profiles of the BCG (red dots) and of the cluster, as traced by the galaxies (blue dots). Error bars are 1 $\sigma$. The green curve and shaded regions are the median and 68\% c.l. predicted by
model 1 (see Table~\ref{t:results} in Sect.~\ref{s:res}). The inset shows the MUSE image of the BCG with the overlaid elliptical annuli used to measure the internal velocity dispersion out to $R\simeq 50$ kpc. {\bf The light-grey circles are sources masked out in the spectral extraction}. }
\label{f:vdp}
\end{figure}

We fitted the BCG surface brightness profile in the $I$ band with a S\'ersic model \citep{Sersic63}, by using {\tt GALFIT} \citep{Peng+11} and the procedure described in \citet{TM23}. The cluster region where the BCG is located is a crowded environment, and it is characterised by the presence of the intra-cluster light. For these reasons, to obtain a robust fit of the 2D galaxy surface brightness of the BCG, we used the methodology described in \citet{TMP18} and \citet{TM23}. This is based on an iterative approach that analyses images of increasing size (to deal with nearest neighbours) and on multiple background estimation (to deal with the intra-cluster light flux contamination). All the pixels belonging to the BCG image, with a flux above the measured background, were considered in the S\'ersic fit. We found a best-fit with an index $n=4.05$, an effective radius $R_e=27.2$ kpc, and a total luminosity $L_{{\rm BCG}} = 4.92 \times 10^{11} \, \lsun$.  We show in Fig.~\ref{f:BCGsersic} that the aperture magnitudes centered on the galaxy are very well reproduced by the Sérsic profile computed with the best-fitting structural parameters. To measure aperture magnitudes we didn't correct for the point-spread-function, and this is the reason why the first point in the Fig.~\ref{f:BCGsersic} is lower that the value expected from the best-fit S\'ersic model and we used a radial range less extended than the region fitted to obtain the 2D surface brightness profile. The S\'ersic index is very close to $n=4$, corresponding to a de~Vaucouleurs profile \citep{deVaucouleurs48}. This is very convenient, since the de-projection of the de~Vaucouleurs profile is well approximated by the Jaffe profile
\citep{Jaffe83},
\begin{equation}
  L_{{\rm Jaffe}}(r) = L_{{\rm BCG}} \, (r/r_J) \, (1+r/r_J)^{-1}. \label{e:jaffe}
\end{equation}
where $r_J=R_e/0.763$. 

The velocity dispersion profile of the BCG has been obtained from \texttt{MUSE} observations (see Fig.~\ref{f:vdp}), by adopting the methodology described in \citealt{Sartoris+20}. We extracted spectra of the BCG in elliptical radial bins from the MUSE data cube, by masking out several interloper galaxies (see light-grey filled circles in the inset of Fig.\ref{f:vdp}). The spatial resolution is limited by the seeing of 0.6\arcsec \citep[see][sampled with 0.2\arcsec pixels]{Caminha+17}, while the velocity resolution is $\sim 10 \, \ks$  for high S/N spectra. The stellar line-of-sight velocity dispersion in each bin is then measured with the pPXF public software \citep{cappellari17}, adopting the same setup parameters and stellar library as in \citet{Sartoris+20}. The precision and accuracy of pPXF in measuring velocity dispersions in different S/N regimes is discussed and tested in \citet{bergamini19}. The six elliptical annuli shown in Fig.~\ref{f:vdp} correspond to radial bins along the semi-major axis, with $a=[0.0-0.6,\, 0.6-1.5,\,1.5-3.0,\,3.0-5.0,\,5.0-8.0,\,8.0-12.0$] arcseconds. The ellipses follow the BCG light with a rotation angle of 185$^\circ$ and axes ratio of $a/b=1.6$.  The circularised radius of each bin is the midpoint of each annulus with equal area, i.e. $R_i=a_{i,mid}\sqrt{b/a}$. Thanks to the 8.5 h long exposure of the MUSE observations, the mean S/N of the spectra extracted in each bin range from 90-100, for the two central bins ($R=1.35,\, 4.72\; {\rm kpc}$), to 21 and 11 for the two outer bins ($R=29.2,\, 44.9\; {\rm kpc}$), respectively. The pPXF cross-correlation procedure not only provides the velocity dispersion in each bin but also the
mean velocity, for which we found a scatter of a few $\ks$, revealing no significant rotational support for the BCG.   

\subsection{The intra-cluster gas}
\label{ss:mgas}
We estimated the gas mass based on dedicated \texttt{Chandra} ACIS-I exposures (ObsId 20544, 209229, 21078, 21079, 21081). We reprocessed them with a standard pipeline based on \texttt{CIAO} 4.10 \citep{Fruscione+06} and \texttt{CALDB} 4.7.8 to create a new events-2 file which includes filtering for grade, status, bad pixels, and time intervals for anomalous background levels. We obtained a cumulative good time interval of 174.0 ksec. All the point sources detected with the \texttt{CIAO} routine \texttt{wavdetect} were masked and not considered in the following analysis. An exposure-corrected image in the 0.7--2 keV band was used to extract a surface brightness profile that was geometrically deprojected (assuming a Galactic absorption $n_H = 3.7 \times 10^{20}$ particles/cm$^2$) to recover an electron density profile $n_e(r)$. A local background 9.4 arcmin far from the X-ray peak, and with no contamination from the cluster emission, was used. The gas mass is then
\begin{equation}
  M_{ICM}(r) = \int_0^r \mu_e \, m_{amu} \, n_e(r)\, 4 \pi x^2 dx,
\end{equation}
where $\mu_e=1.16$ is the mean electron mass weight appropriate for a fully ionized plasma with 30\% solar abundances \citep{agss09}, and $m_{amu} = 1.66 \times 10^{-24}$ g is the atomic mass unit.

A full spectral analysis provided the measurements of the gas temperature in azimuthally averaged bins up to $\sim$800 kpc. By following the backward approach to reconstruct the hydrostatic mass as described in \cite{Ettori+10}, we constrained the total mass profile using a NFW model with best-fit parameters $c_{200}=2.8^{+1.1}_{-0.9}$ and $R_{200} = 2.2^{+0.4}_{-0.3}$ Mpc.

\section{The dynamical analysis: method} \label{s:method}
To determine the cluster mass profile we performed a simultaneous maximum likelihood fit to the BCG velocity dispersion profile, and to the velocity distribution of cluster members as a function of radius, out to 2.2 Mpc. This radius corresponds to a 2 $\sigma$ upper limit to the virial radius $\rtwo$\footnote{Here and throughout this paper, we call $\rv$ the radius that encloses an average density $\Delta$ times the critical density at the halo redshift. $\mv$ is related to $\rv$ by  $\mv \equiv \Delta/2 \, \rm{H_z} \, \rv^3$/G, where $\rm{H_z}$ is the Hubble constant at the cluster redshift and G is the gravitational constant.}  
estimate of the cluster from the gravitational lensing analysis of \citet{Umetsu+12}. We used an extension of the \texttt{MAMPOSSt} method\footnote{The new method is described in \citet{Pizzuti+23} and freely available at \texttt{https://github.com/Pizzuti92/MG-MAMPOSSt} and on Zenodo \citep{lorenzo_pizzuti_2023_7515195}.} 
originally developed by \citet{MBB13}, that we already used in the dynamical analysis of the AS1063 cluster \citep{Sartoris+20}.
In this extension of the original code, \texttt{MAMPOSSt} estimates the likelihood ${\cal L}_{{\rm gal}}$ of the observed projected phase-space distribution of cluster members, and combines this likelihood with the likelihood resulting from the $\chi^2$ fit to the observed, line-of-sight (l.o.s. in the following) BCG stellar velocity dispersion profile,
\citep[using eqs.~(9) and (26) in][]{MBB13}. The inputs to \texttt{MAMPOSSt} are the individual cluster-centric radial distances and rest-frame velocities of the cluster members (identified as described in Sect.~\ref{ss:members}), and the surface brightness and velocity dispersion profiles of the BCG in radial concentric bins. 

The final likelihood of the model is given by the sum $-\ln {\cal L}_{{\rm tot}}=-\ln {\cal L}_{{\rm gal}}+\chi^2_{\rm{BCG}}/2$
\citep[based on the theorem of][]{Wilks38}. The following profiles enter the
${\cal L}_{{\rm tot}}$ determination:
\begin{enumerate}
    \item the number density profile of the BCG stars, $\nu_{{\rm BCG}}(r)$,
    \item the number density profile of the cluster members, $\nu_{{\rm gal}}(r)$,
    \item the cluster total mass profile, $M(r)$,
    \item the velocity anisotropy profile of the BCG stars, $\beta_{{\rm BCG}}(r)$,
    \item the velocity anisotropy profile of the cluster galaxies, $\beta_{{\rm gal}}(r)$,
\end{enumerate}
where $\beta \equiv 1-(\sigma_{\theta}/\sigma_r)^{2}$ and $\sigma_{\theta}$ and $\sigma_r$ are the tangential and radial component of the velocity dispersion, respectively. In the following we describe how we modeled these profiles for our dynamical analysis.

\subsection{The number density profiles} \label{ss:nu}
We assumed $\nu_{{\rm BCG}}(r)$ to coincide with the de-projection of the BCG surface brightness profile, that we approximated with a Jaffe profile \citep[][see eq.~\ref{e:jaffe}]{Jaffe83}.

\begin{figure}[ht]
\includegraphics[width=0.5\textwidth]{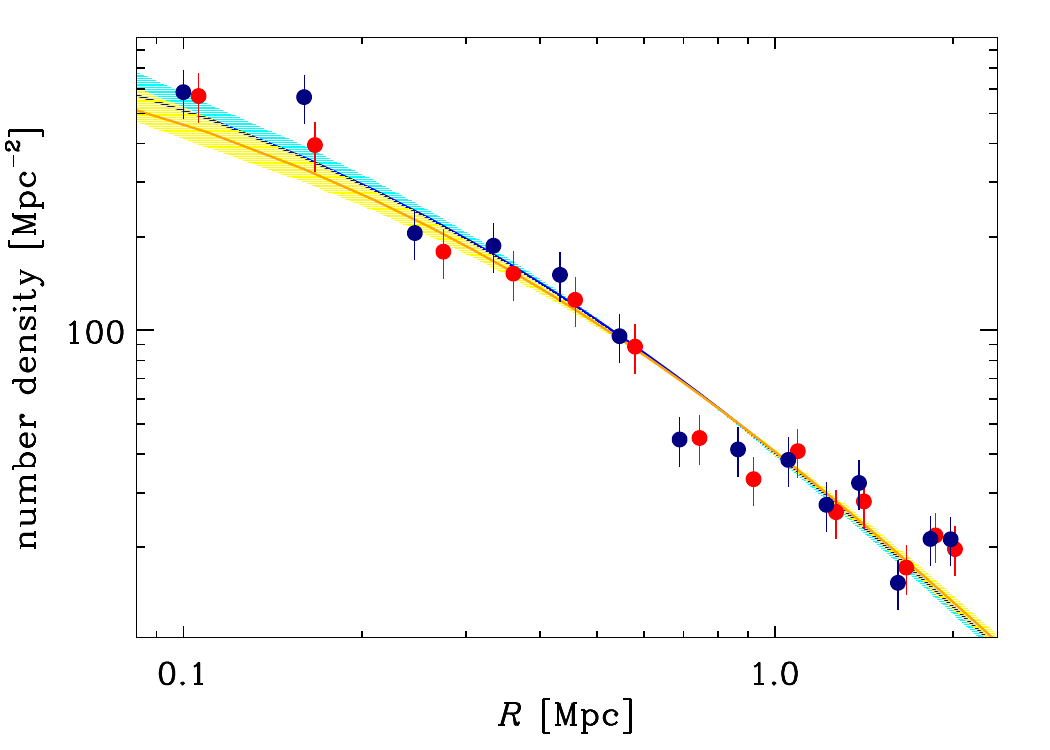}
\caption{The projected number density profiles of cluster members (dots with 1 $\sigma$ error bars), corrected for spectroscopic incompleteness, and their best-fits with projected NFW models (lines with 1 $\sigma$ confidence regions). \texttt{CLUMPS} sample: red points, orange line and yellow confidence region. \texttt{P+G} sample: blue navy points, blue line and cyan region. }
\label{f:denprofs}
\end{figure}

To determine $\nu_{{\rm gal}}(r)$ we had to first account for the incompleteness of our spectroscopic data set. What is relevant here is only the relative completeness at different cluster-centric radii. Our \texttt{MUSE} observations are much deeper than our \texttt{VIMOS} observations, but are restricted to a very central region. To reduce the difference in spectroscopic completeness between the region covered by \texttt{MUSE} observations and other regions, and limit the amount of completeness correction, in the determination of $\nu_{{\rm gal}}(r)$ we restricted the samples of members to a $R_{\rm C}$-band magnitude $\leq 24$, that is approximately the limit reached by the non-\texttt{MUSE} redshift determinations. This magnitude cut leaves $\sim 90$\% of the members in the full sample (433 and 439 galaxies within 2.2 Mpc in the \texttt{CLUMPS} and \texttt{P+G} samples, respectively), that can be considered representative of the full sample. We estimated the spectroscopic completeness as a function of cluster-centric radii, following the procedure of \citet{Biviano+13}. The region covered by \texttt{MUSE} observations is complete to the chosen magnitude limit, while the remaining subsample has a radially decreasing completeness, from a value of $\sim 0.8$ near the center to $\sim 0.7$ at the virial radius.

We used a maximum likelihood technique \citep{Sarazin80} to fit a projected NFW model \citep{Bartelmann96} to the projected number density profile of cluster members, separately for the \texttt{CLUMPS} and \texttt{P+G} samples. In the fitting procedure, we weighted each galaxy by the inverse of its radial completeness. The best fits are shown in Fig.~\ref{f:denprofs}. There is only one free parameter in the fit, the scale radius $\rnu$ of the NFW model, since the normalization of the fitting function is constrained by the requirement that the total (completeness corrected) number of observed galaxies is identical to the integral of the fitting function over the radial range of the fit.  By definition, the best fit $\rnu$ values of the projected NFW profiles are the same also for the 3D NFW profiles. We found very similar best fit values  for the \texttt{CLUMPS} and \texttt{P+G} sample, $\rnu=0.46_{-0.07}^{+0.08}$ and $0.40_{-0.06}^{+0.07}$ Mpc, respectively. These values are consistent within 1 $\sigma$ with the estimate of \citet[][$0.63_{-0.09}^{+0.11}$]{Biviano+13}, albeit slightly smaller, but they did not have \texttt{MUSE} data, they cut the sample one magnitude brighter than we do, and they ignored the 20\% change in radial completeness of their sample. 

\subsection{The mass profile} \label{ss:mass}
We model the total cluster mass profile as the sum of several components,
\begin{equation}
  M(r) = M_{{\rm DM}}(r) + M_{{\rm BCG}}(r) +  M_{{\rm gal}}(r)+ M_{{\rm ICM}}(r), \label{e:mtgg}
\end{equation}
where $M_{{\rm DM}}$ is the DM profile, while 
$M_{{\rm BCG}}, M_{{\rm gal}}$, and $M_{{\rm ICM}}$ are the baryonic mass profiles, namely
the BCG stellar mass profile, the stellar mass of all other member galaxies ('satellites' hereafter), and the mass of the hot intra-cluster gas. We characterize the DM profile by a gNFW model,
\begin{eqnarray}
  M_{{\rm gNFW}}(r) = & \mtwo \, (r/r_{200})^{3-\gdm}  \\
  & \times \frac{_{2}F_{1}(3-\gdm,3-\gdm,4-\gdm,-r/\rs)}{_{2}F_{1}(3-\gdm,3-\gdm,4-\gdm,-r_{200}/\rs)},  \nonumber
  \label{e:mgnfw}
\end{eqnarray}
where $_{2}F_{1}$ is the hypergeometric function \citep{Mamon+19}.
There are three free parameters in this model, $\rtwo$, $\rs$, and $\gdm$. The BCG stellar mass profile is given by $M_{{\rm BCG}}(r) = (M/L) \, L_{{\rm Jaffe}}(r)$ (see eq.~\ref{e:jaffe}) and the only free parameter is the mass-to-light ratio $M/L$. We take the satellite stellar mass profile $M_{{\rm gal}}(r)$ from \citet{Annunziatella+14}. The hot intra-cluster gas mass profile $M_{{\rm ICM}}(r)$ is derived as described in Sect.~\ref{ss:mgas}.
  We consider $M_{{\rm gal}}$ and $M_{{\rm ICM}}$ as fixed and neglect their uncertainties, since their contribution to the total mass near the cluster center is very marginal (see Fig.~\ref{f:mtot}).

\subsection{The velocity anisotropy profiles} \label{ss:beta}
\citet{Kronawitter+00} found a remarkable homogeneity in the velocity anisotropy profile $\beta(r)$ of giant ellipticals. The studies of \citet{Kronawitter+00} and \citet{Santucci+23} indicate that giant ellipticals have at most mild velocity anisotropy, $\beta \lesssim 0.3$, . We therefore try two constant $\beta$ model, namely $\beta=0$ and $\beta=0.3$ at all radii. To assess the impact of our assumption we also consider a third model, the OM model \citep{Osipkov79,Merritt85-df} in which $\beta(r)$ rises fast with radius,
\begin{equation}
  \beta_{{\rm BCG}}(r)= r^2 / (r^2+r_{\beta}^2),
\end{equation}
where we assume $r_{\beta}=r_{J}$. 

The velocity dispersion profile of galaxy clusters is less well constrained than that of the BCG, and probably much less homoegenous across different clusters \citep[see, e.g.,][]{WL10,MBB13,Biviano+13,Munari+13,Mamon+19}. Lacking strong priors on the cluster $\beta(r)$ we adopt a very generic model with two free parameters indicating the anisotropy at $r=0$, $\beta_0$, and at large radii, $\beta_{\infty}$ \citep{Tiret+07},
\begin{equation}
  \beta_{{\rm gal}}(r)=\beta_0+(\beta_{\infty}-\beta_0) \, r/(r+r_2).
\end{equation}
In our analysis we provide constraints on $\beta_0$ and $\beta_{\infty}$ in terms of the equivalent parameters $\mbox{$(\sigma_r/\sigma_{\theta})_0=(1-\beta_0)^{-1/2}$}$ and $\mbox{$(\sigma_r/\sigma_{\theta})_{\infty}=(1-\beta_{\infty})^{-1/2}$}$.

\section{The dynamical analysis: results} \label{s:res}
\begin{figure*}[ht]
\includegraphics[width=1.\textwidth]{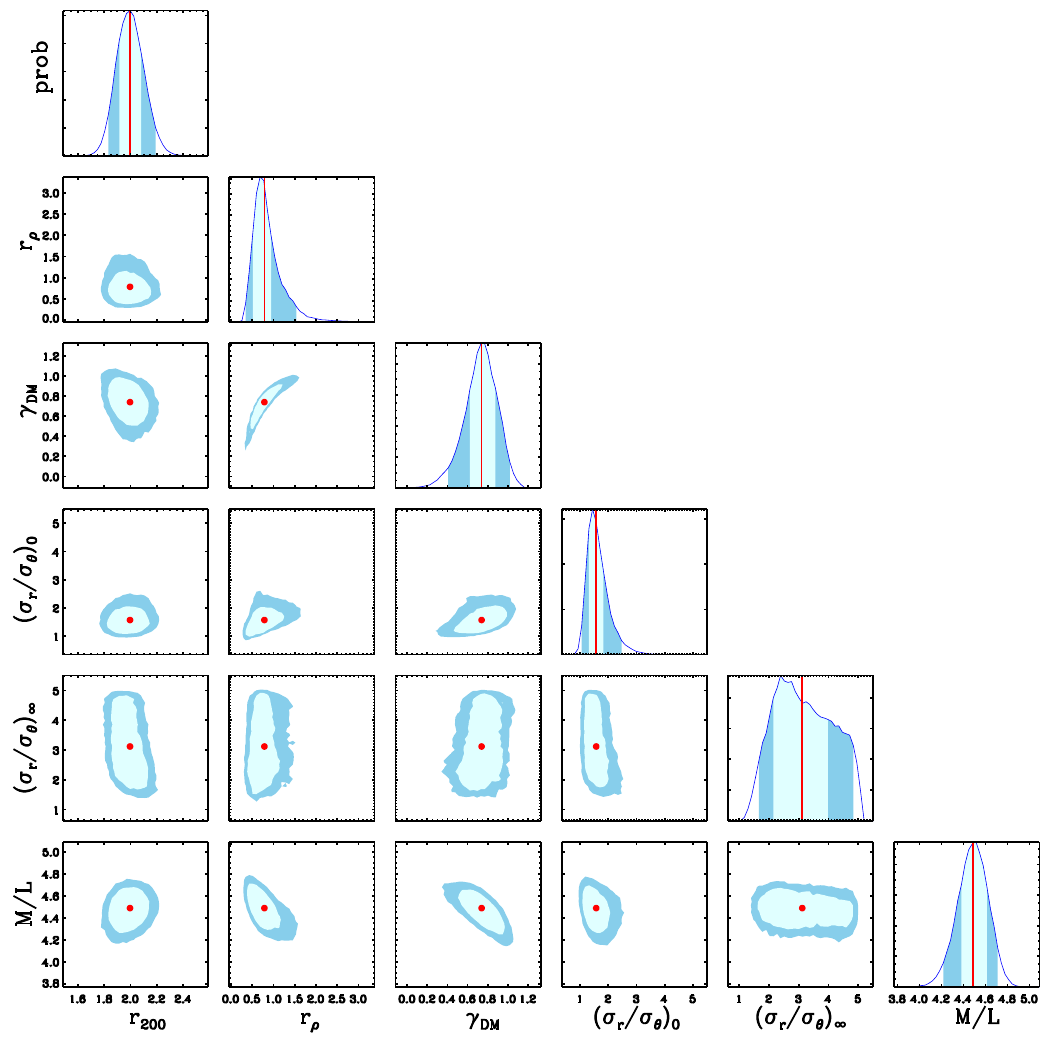}
\caption{Result of the \texttt{MAMPOSSt} analysis on the \texttt{CLUMPS} sample, using $\beta_{{\rm BCG}}=0.0$ (model 1), based on the marginalization of the posterior distribution resulting from the 100,000 samples of the MCMC analysis. The red lines and dots indicate the median of the distribution, The light (resp. dark) blue regions indicate the 68\% (resp. 95\%) c.l.. The BCG $M/L$ is relative to the $I$-band luminosity.}
\label{f:triplot}
\end{figure*}

We determine the marginal distributions of the free parameters in the dynamical analysis by
 a Monte Carlo Markov Chain (MCMC) sampling of 100,000 points in the parameter space. The free parameters in our \texttt{MAMPOSSt} runs are the following: $\rtwo, \rs, \gdm, M/L, (\sigma_r/\sigma_{\theta})_0, (\sigma_r/\sigma_{\theta})_{\infty}$, plus the $\rnu$ parameter for which we assume the $\pm 1 \sigma$ interval derived externally (see Sect.~\ref{ss:nu}) as a flat prior in the \texttt{MAMPOSSt} runs. 
We run \texttt{MAMPOSSt} on both our  fiducial sample of cluster members, based on the \texttt{CLUMPS} algorithm, and, for comparison, also on our \texttt{P+G} sample of cluster members,
using all three BCG velocity anisotropy models described in Sect.~\ref{ss:beta}.  In Table~\ref{t:results}, we list the median values and 68\% confidence levels (c.l. hereafter) of all parameters, for the \texttt{CLUMPS} sample and the three $\beta_{{\rm BCG}}$ models (numbered 1 to 3 in the Table), as well as for the \texttt{P+G} sample with  $\beta_{{\rm BCG}}=0.0$ (model 4 in the Table; we do not show the results for the other BCG anisotropy profiles for the sake of conciseness). The results for model 1 are displayed in the triangular plot of Fig.~\ref{f:triplot}.  

\begin{table*}
\begin{center}
\caption{Results of the \texttt{MAMPOSSt} analysis}
\label{t:results}
\begin{tabular}{cccccccccc}
\hline
Model id. & Sample & $\beta_{{\rm BCG}}$ & $\rtwo$ & $\rs$ & $\gdm$ & $M/L$ & $(\sigma_r/\sigma_{\theta})_0$ & $(\sigma_r/\sigma_{\theta})_{\infty}$ & $\Delta \mathrm{BIC}$ \\
& & & (Mpc) & (Mpc) & & ($M_{\odot}/L_{\odot}$)  & & & \\
& & & \multicolumn{6}{c}{[68 \% confidence levels]} & \\
& & & \multicolumn{6}{c}{[95 \% confidence levels]} &  \\
\hline
1 & \texttt{CLUMPS} & 0.0 & 2.00  & 0.87 & 0.73  & 4.48 & 1.65  & 3.19  &  0.9  \\
& &  & [1.89-2.10] &  [0.48-1.02] & [0.59-0.91] & [4.36-4.63] & [1.20-1.87] & [2.04-4.09] &  \\
& &  & [1.81-2.21] &  [0.33-1.60] & [0.39-1.04] & [4.20-4.74] & [1.00-2.48] & [1.59-4.91] & \\
2 & \texttt{CLUMPS} & 0.3 & 1.99  & 1.01 & 0.76  & 3.91  & 1.81  & 3.17  &  0.0 \\
& & & [1.88-2.09] & [0.50-1.20] & [0.61-0.95] & [3.77-4.07] & [1.31-1.98] & [2.10-4.16] &  \\
& & & [1.79-2.22] & [0.30-2.03] & [0.39-1.10] & [3.59-4.19] & [1.08-2.83] & [1.48-4.92] & \\
3 & \texttt{CLUMPS} & OM  & 1.96  & 0.98  & 0.21  & 4.64  & 2.07  & 3.48 & 13.6  \\
& & & [1.84-2.06] & [0.69-1.04] & [0.01-0.28] & [4.56-4.72] & [1.45-2.26] & [2.59-4.53] &  \\
& & & [1.74-2.19] & [0.59-1.58] & [0.00-0.53] & [4.48-4.79] & [1.24-3.26] & [1.89-5.03] & \\
4 & \texttt{P+G} & 0.0    & 1.97  & 0.83 & 0.71 & 4.49  & 1.60  & 2.91 & 17.9  \\
& & & [1.84-2.08] & [0.44-0.98] & [0.55-0.91] & [4.36-4.64] & [1.16-1.71] & [1.59-3.81]  &  \\ 
& & & [1.75-2.21] & [0.27-1.67] & [0.33-1.06] & [4.19-4.64] & [0.95-2.44] & [1.15-4.88] \\ [0.10cm]
\hline
\end{tabular}
\end{center}
\tablecomments{For each model we 
provide in the first row the median values of the  parameters from the marginal distribution obtained with the \texttt{MAMPOSSt} analysis, in the second, and, respectively, third row, the 68\% and 95\% c.l. on these parameters. The BCG $M/L$ is relative to the $I$-band luminosity.
The last column gives the value of $\Delta \mathrm{BIC}$ with respect to the best fit model 2
resulting from the comparison between the observed velocity dispersion profiles of the BCG and the cluster and those predicted by the best fit \texttt{MAMPOSSt} models. }
\end{table*}

The best-fit parameter values do not depend on the sample, but some of them do depend on the choice of BCG velocity anisotropy models, even if most differences are within the 95\% c.l.. In particular,
\begin{itemize}
    \item $\gdm$ is higher for the two $\beta_{{\rm BCG}}$ constant models (1 and 2 in Table~\ref{t:results}) than for model 3 that uses the OM $\beta_{{\rm BCG}}(r)$;
    \item the $I$-band BCG $M/L$ is lowest for model 2 ($\beta_{{\rm BCG}}=0.3$). 
\end{itemize}
To select among the different models, we evaluate how well they fit the observed line-of-sight (l.o.s. hereafter) velocity dispersion profiles of the BCG and the cluster. We compute the model-predicted velocity dispersion profiles by applying eqs.~(9) and (26) in \citet{MBB13}. We then evaluate 
\begin{equation}
    \chi^2 = \sum_{i=1}^n (\sigma_{i,{\rm M}}-\sigma_{i,{\rm o}})^2/\delta_{i,{\rm o}}^2,
\label{e:chi2}
\end{equation}
where $\sigma_{i,{\rm o}}$ and $\sigma_{i,{\rm M}}$ are, respectively, the observed velocity dispersion and the average of the profiles resulting from the 100,000 parameter samples of the MCMC analysis, and $\delta_{i,{\rm o}}$ is the 1 $\sigma$ error on the observed velocity dispersion, all quantities evaluated for the $i$-th of $n$ radial bins. We then use the Bayes Information Criterion \citep[BIC,][]{Schwarz78}, to compare the quality of the fits of the different models, $\Delta \mathrm{BIC}=\Delta \chi^2 -k \Delta \ln N$, where $k$ is the number of free parameters and $N$ the number of data, which is slightly different for model 4 and the other models. the $\Delta \mathrm{BIC}$ values are listed in Table~\ref{t:results} with respect to the minimum value of $\chi^2$ among the four models.

According to \citet{KR95}, a value $\Delta \mathrm{BIC} > 10$ indicates strong evidence against the  model with the largest BIC, while a value $\Delta \mathrm{BIC} < 2$ does not indicate a significant difference in the quality of the fits of the two models compared. We then conclude that
the $\beta_{{\rm BCG}}=0.0$ Model 1 and $\beta_{{\rm BCG}}=0.3$ Model 2 provide an equally good fit to the data,
and a significantly better fit than the OM $\beta_{{\rm BCG}}$ Model 3, 
for both the \texttt{CLUMPS} and the \texttt{P+G} samples.
The fits obtained using the \texttt{P+G} sample are significantly worse than those obtained using the \texttt{CLUMPS} sample.
This is due to the inclusion of a probable foreground group of galaxies in the \texttt{P+G} sample of members, that increases the l.o.s. velocity dispersion of the cluster near the center (see Fig.~\ref{f:members}). Given the results of the $\chi^2$ analysis, in the following we only consider Models 1 and 2, both based on the \texttt{CLUMPS} sample. 

Since there is tension in the value of the $M/L$ parameter obtained by using the $\beta_{{\rm BCG}}=0.0$ and 0.3 Models 1 and 2, we look for external constraints on $M/L$. 
\citet{CvD12} estimated the $M/L$ of 38 ellipticals using stellar population synthesis models based on a variable initial mass function. They found that the velocity dispersion of low-$z$ early-type galaxies correlates with their $M/L$. Using the data of Table~2 in their paper, we fit the relation
$M/L = -0.4 + 1.6 \, \sigma_{{\rm BCG}}$,
where the BCG luminosity is observed in the $I$-band, and
$\sigma_{{\rm BCG}}$ is the observed BCG velocity dispersion in units of $100 \, \ks$, measured within an aperture of $R_e/8$. To apply this relation to the BCG of MACS~1206, we apply the evolutionary and K- corrections
to the $I$-band magnitude of our BCG, using the tables of \citet{Poggianti97}, updated for the cosmology used in this paper. We find that the two corrections almost cancel out, giving a $\sim 4$\% difference in luminosity that we ignore in this consideration. The MACS~1206 BCG has $\sigma=315$ km~s$^{-1}$ within the radius $R_e/8=3.4$ kpc, that would indicate $M/L=4.6 \, M_{\odot}/L_{\odot}$. This value is marginally closer to the solution we found for the $\beta_{{\rm BCG}}=0.0$ model 1, which we therefore consider our reference model for the following discussion. Our
model 1 dynamically-derived M/L value supports \citet{CvD12}'s conclusion of a bottom-heavy IMF for high-velocity dispersion ellipticals.

 The predicted BCG and cluster velocity dispersion profiles for Model 1 is compared to the observed profiles in Fig.~\ref{f:vdp}. Note that the BCG velocity dispersion profile does not join onto that of the cluster galaxies, and this is mostly due to the fact that the BCG stars and galaxies do not follow the same spatial distribution \citep[i.e. the BCG and galaxies $\nu(r)$ in eq.~(9) of][are different]{MBB13}, and partly due to their different orbital distributions. Overall, the fit appears very good, although the formal value of the $\chi^2$ is rather high, $30.8$ for 14 data points and 7 free parameters (it is only slightly smaller for Model 2), due to the very 
 small error bars of the velocity dispersion profile of the BCG. 
\begin{figure}[ht]
\includegraphics[width=.45\textwidth]{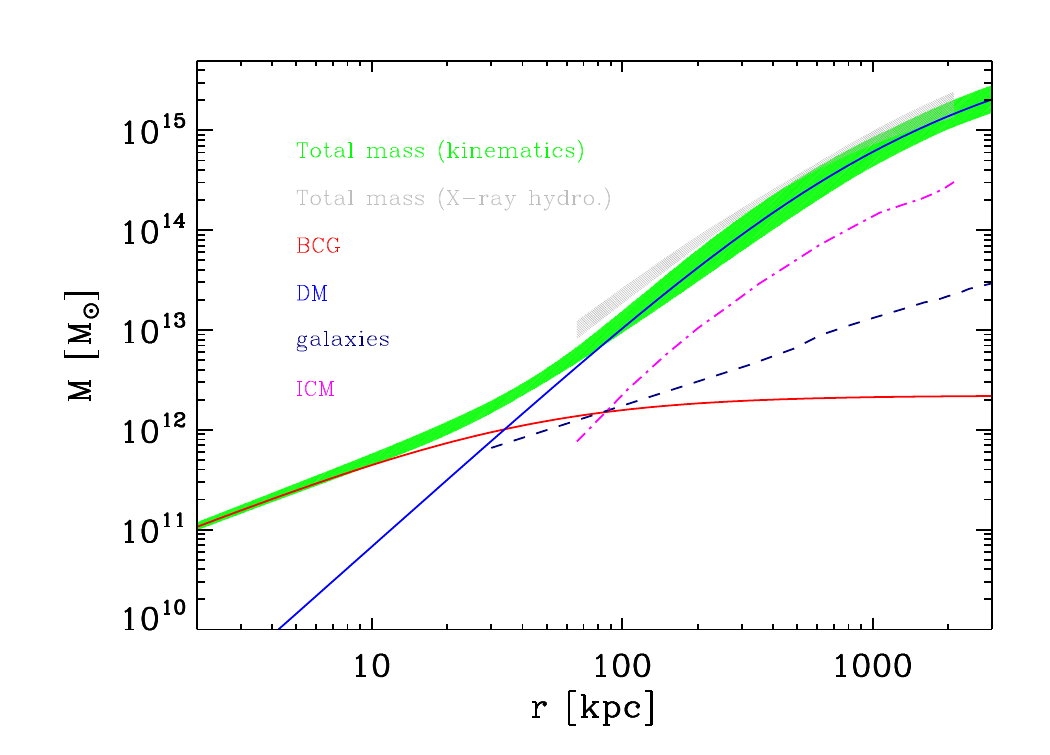}
\caption{Total mass profiles $M(r)$ of the MACS~1206 cluster. Green (grey) shading: 68\% confidence region for the total mass profile obtained from the kinematical analysis -- Model 1 (respectively: from the X-ray hydrostatic analysis). Blue (resp. red) solid line: DM profile (resp. BCG stellar mass profile). Navy blue dashed line: satellites stellar mass profile. Magenta dash-dotted line: intra-cluster gas mass profile. }
\label{f:mtot}
\end{figure}

In summary, Model 1 is characterized by a $M(r)$ with a DM inner slope slightly shallower than the NFW profile, isotropic stellar orbits of the BCG, and radially elongated orbits of the cluster galaxies, with increasing anisotropy at larger radii. The increasing anisotropy with radius is a feature seen in cluster-size halos from cosmological simulations \citep[see, e.g.][]{Munari+13,Lotz+19}, and also found by \citet{Biviano+13} for the same cluster, but
Model 1 predicts a stronger radial anisotropy, especially near the center,  although the difference is less than 95\% significant. The total $M(r)$ of Model 1 and its components are shown in Fig.~\ref{f:mtot}. 

In Fig.~\ref{f:mtot} we also show the total $M(r)$ obtained by the X-ray hydrostatic analysis (see Sect.~\ref{ss:mgas}). This $M(r)$ is consistent with the $M(r)$ we obtain from kinematics at radii $\gtrsim 300$ kpc, but lies above it at smaller radii. The difference can be attributed to the fact that we adopted the NFW model in the X-ray analysis, since the X-ray data alone do not allow constraining the inner slope of a more generic gNFW model. Cluster masses derived from X-ray data are often claimed to be affected by the so-called hydrostatic bias \citep[e.g.][]{LKN09,Rasia+06,Hoekstra+15}. Our analysis does not show significant evidence of the hydrostatic bias in this cluster, in agreement with our previous findings in Abell~S1063 \citep{Sartoris+20}. In particular, the X-ray hydrostatic analysis does not appear to under-estimate the cluster mass.

\begin{figure}[ht]
\includegraphics[width=.5\textwidth]{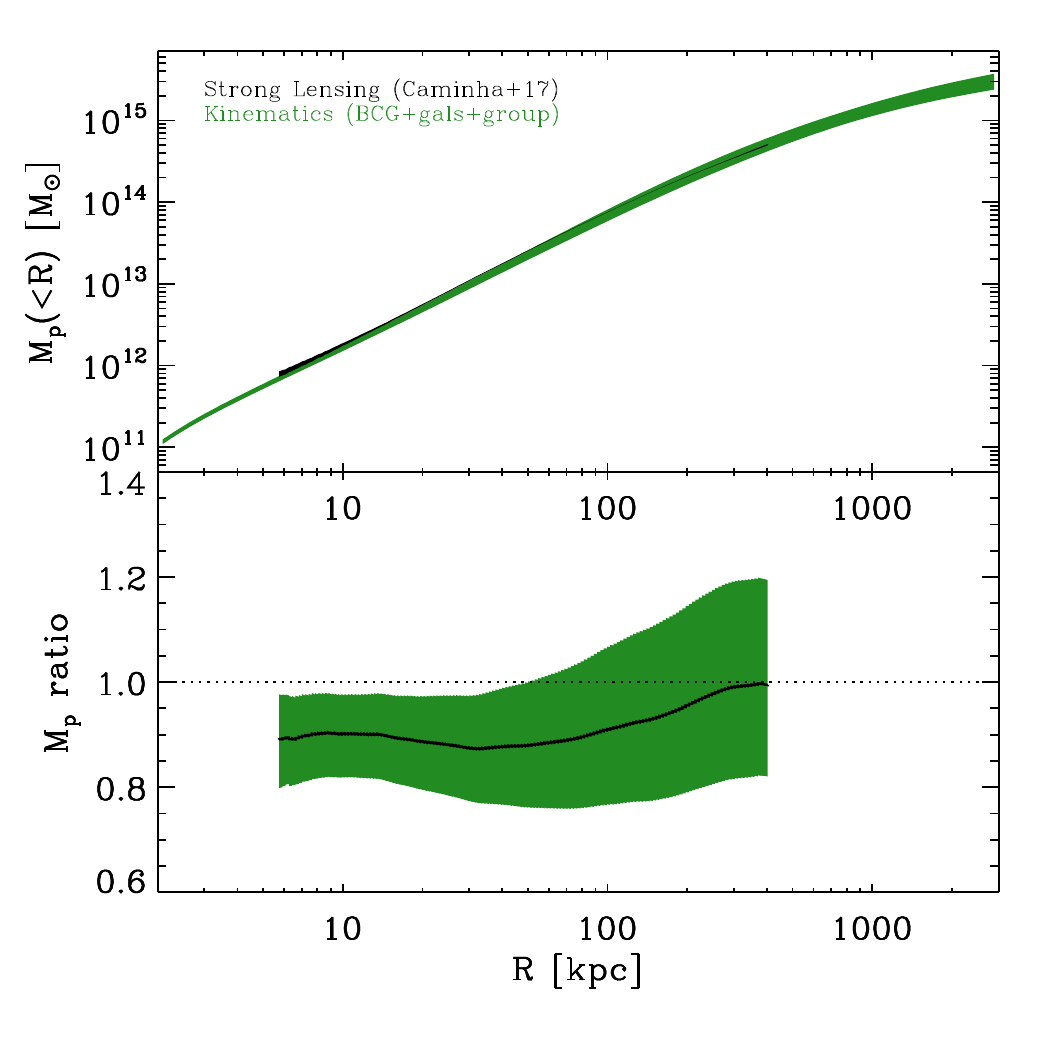}
\caption{{\it Top panel:} projected mass profiles $M_p(<R)$ of the MACS~1206 cluster. Shadings indicate the 68\% confidence regions on $\mpr$. Green shadings: $\mpr$ obtained from the kinematics analysis described in this paper, with 
the contribution of a foreground group. 
Black shading: $\mpr$ obtained from the strong lensing analysis of \citet{Caminha+17}. 
{\it Bottom panel:} ratio of kinematics to strong lensing
$M_p(<R)$. 
Shadings indicate 68\% c.l.
}
\label{f:mp}
\end{figure}

We compare the mass profile from Model 1 to the mass profile obtained by \citet{Caminha+17} based on a strong lensing analysis.
We emphasize that the unusually large number of inner radial multiple images in M1206 (see Fig.2 of \citealt{Caminha+17}) translates into a very tight constraint of the (projected) inner density profile slope down to $R\simeq 50\,{\rm kpc}$.
We show this comparison in Fig.~\ref{f:mp}, where we display the projected mass profiles $\mpr$, rather than the 3D ones, to ease the comparison with the strong lensing $\mpr$ of \citet{Caminha+17}. 
To allow for a fair comparison between the kinematics and lensing estimates we must take into account the mass contribution by the foreground group already discussed in Sect.~\ref{ss:members}. This contribution is naturally included in the lensing estimate, but not in the estimate from kinematics, since the group galaxies are correctly excluded from our dynamical analysis since they are not cluster members.

This group of 15 galaxies is located close to the cluster center ($<200$ kpc) and has a l.o.s. velocity dispersion of $484_{-89}^{+107}$ km~s$^{-1}$. Using the "AGN gal" relation of Table~1 in \citet{Munari+13}, we estimate a group mass $\mtwo \sim 4-17 \times 10^{13} \, \msun$. This group was already identified by \citet{Young+15} as residual Sunyaev-Zel'dovich emission left after subtracting a model emission by the main cluster. \citet{Young+15} also estimated the group mass by several methods. Our mass estimate is consistent with theirs \citep[see Table~7 in][]{Young+15}.

We do not have sufficient spectroscopic data to constrain the group mass concentration, so we use the mass-concentration relations of \citet{DM14} and \citet{Correa+15}, to estimate $\ctwo \sim 5$.   Assuming that the group mass distribution follows a NFW profile, and identifying the group center with the cluster center for simplicity, we can then estimate the group $M_p(<R)$ and add it to the cluster $M_p(<R)$ obtained from kinematics. We thus obtain the profile shown by the dark green shading in Fig.~\ref{f:mp}, 
that is within 68\% c.l. of the strong lensing $\mpr$ for $R > 50$ kpc, and only slightly below at smaller radii. The remaining $\sim 10$\% disagreement may be due to our simplified assumptions on the mass profile of the foreground group and the position of its center, as well as on the spherical symmetry assumption used to project the cluster and group mass profiles.

\section{Discussion}\label{s:disc}
\begin{figure}[ht]
\begin{center}
\includegraphics[width=.5\textwidth]{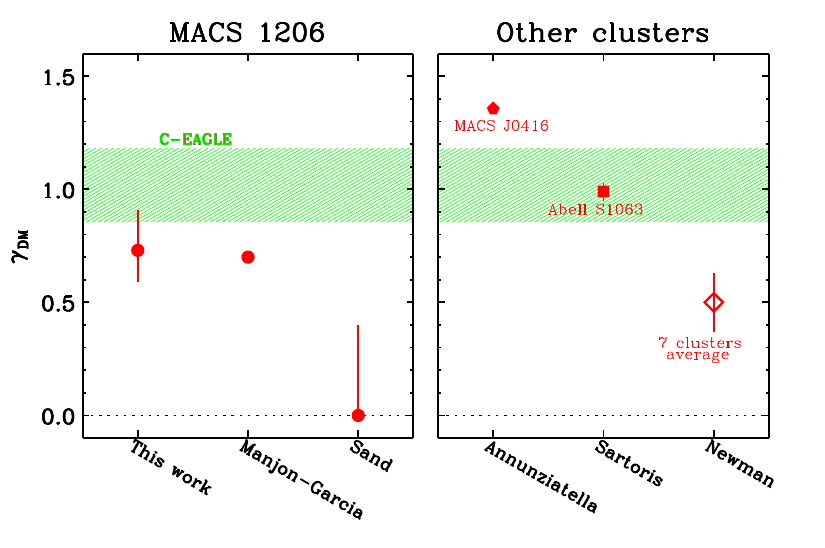}
\end{center}
\caption{Summary of $\gdm$ measurements for clusters of galaxies.
Error bars are 1 $\sigma$. The green region indicates the standard deviation around the mean value for 16 massive halos in the \texttt{C-EAGLE} simulations
analysed by \citet{He+20}. {\it Left panel:} results for MACS~1206. No error estimate is available for the value of \citet{MGDHL20}.
{\it Right panel:} results for other clusters, MACS J0416$-$2403 \citep[pentagon,][]{Annunziatella+17}, Abell~S1063 \citep[square,][]{Sartoris+20}, and an average of seven cluster values \citep[open diamond,][]{Newman+13b}. The error bar for the value of \citet{Annunziatella+17} is smaller than the symbol size, but not fully comparable to the other error bars in the figure, because it is based on the simplified assumption of a power-law mass density profile.}
\label{f:gammas}
\end{figure}

Our result for the inner slope of the DM density profile is $\gdm=0.73_{-0.14}^{+0.18}$ (68\% c.l.), with an upper 95\% c.l. of 1.04, consistent with the NFW slope (see Table~\ref{t:results}). Previous results on the inner slope of the MACS~1206 mass density profile were obtained by \citet{Umetsu+12}, \citet{Young+15}, and \citet{Caminha+17} for the {\em total} mass and by \citetalias{Sand+04} and \citet{MGDHL20} for the DM; we show the latter two results in Fig.~\ref{f:gammas} together with our own. \citet{Umetsu+12} combined a weak-lensing distortion, magnification, and strong-lensing analysis, and found an inner slope value $\gamma_{{\rm tot}}=0.96_{-0.49}^{+0.31}$ (68\% c.l.). An almost identical, but more precise result, was obtained by the strong lensing analysis of \citet{Caminha+17}, who found $\gamma_{{\rm tot}}=0.91 \pm 0.04$ (68\% c.l.). \citet{Young+15} modeled the Sunyaev-Zel'dovich emission by the cluster with a gNFW profile with $\gamma_{{\rm tot}}=0.7$ (no error bars provided). The results of \citet{Umetsu+12}, \citet{Young+15}, and \citet{Caminha+17} are not directly comparable to ours, as they did not constrain the DM profile inner slope. 

According to \citet{Schaller+15}'s analysis of the \texttt{EAGLE} cosmological simulations \citep{Crain+15,Schaye+15}, very massive halos/clusters (such as MACS~1206) have $\gtot-\gdm \approx 0.1$, on average, a difference attributed to the stellar mass contribution of the BCG at the cluster center. If clusters have a similar baryon-to-total mass distribution as the numerical halos in the \texttt{EAGLE} simulations, we can conclude that the results of \citet{Umetsu+12} and \citet{Caminha+17} for $\gtot$, are fully consistent with our result for $\gdm$ \citep[we cannot make any statement about the result of][because we ignore its uncertainty]{Young+15}. However, our result for the BCG $M/L$ supports a bottom-heavy IMF, while the \texttt{EAGLE} simulations adopted a more conventional \citet{Chabrier03} IMF \citep{Schaller+15} to convert the stellar masses to luminosities. The difference in the IMF implied by our dynamical analysis and the IMF adopted in the \texttt{EAGLE} simulations could be compensated by the fact that the simulated BCGs contain more stellar mass than observed BCGs by up to 0.6 dex \citep{He+20}. Given these uncertainties it is not straightforward to estimate $\gdm$ from the observed $\gtot$ values.

\citet{MGDHL20} determined the MACS~1206 DM density profile via a strong lensing analysis, by subtracting the BCG stellar mass distribution. Unlike us, they did not account for the mass contributions of the satellites and the hot ICM, but these are probably negligible in the inner cluster region. They found $\gdm=0.7$, which is in excellent agreement with our result, but they did not provide the uncertainty in their measurement. 

\citetalias{Sand+04} combined the dynamical analysis of the BCG stellar velocity dispersion with the constraints obtained by modelling the giant gravitational arcs, to determine the inner slope of the DM density profile. Also in this case, the satellites and ICM mass contributions were neglected.
They found $\gdm=0$, with a 68\% (respectively 95\%) upper limit $\gdm \leq 0.40$ (respectively, $0.67$). Their result is therefore inconsistent with the NFW slope at the 95\% c.l., and with our own result at the 68\% c.l.. A cored inner mass density profile was also found to be favored over a cuspy model by \citet{LBJ22}, even if these authors did not attempt to provide an estimate of $\gdm$.

Part of the difference in \citetalias{Sand+04}'s $\gdm$ value and ours may be related to the difference in the values of $\rs$, since there is a strong covariance between $\rs$ and $\gdm$, as shown in Fig.~\ref{f:triplot} and as already pointed out by \citet{He+20} in a more general case. Our large spectroscopic data set for cluster galaxies allows us to directly constrain $\rs$. On the other hand, \citetalias{Sand+04} were unable to constrain $\rs$ from their lensing data and they had to fix it to an {\em ad hoc} value, $\rs=0.4$ Mpc. Their adopted value agrees with the estimate of \citet[][$0.28 \pm 0.06$ Mpc for a gNFW model]{Umetsu+12} based on gravitational lensing, but not with the estimate of \citet[][$0.64 \pm 0.15$ Mpc]{Donahue+14} based on X-ray Chandra data. Our independent analysis of X-ray data confirms the value of \citet{Donahue+14}; we find $\rs=0.80_{-0.26}^{+0.35}$ Mpc in excellent agreement with the value we obtain from our kinematical analysis\footnote{Different values of $\rs$ can also explain why the velocity anisotropy profile $\beta(r)$ of the cluster galaxies we find is more than 1~$\sigma$ above the one found by \citet[][see their Fig.~15]{Biviano+13}, who solved the inverse Jeans equation by assuming the \citet{Umetsu+12} $M(r)$ in the virial region.} (see Table~\ref{t:results}). Maybe the smaller $\rs$ value obtained by the lensing analyses compared to the X-ray and kinematics analyses, is due to the projected group near the center that we discussed at the end of Sect.~\ref{s:res}. In any case, even if we force $\rs$ to the value adopted by \citetalias{Sand+04} we would find $\gdm \approx 0.6$, still significantly greater than zero
(see Fig.~\ref{f:triplot}). 

We think that the main reason for the difference between our and \citetalias{Sand+04}'s $\gdm$ values is in the BCG data. While the BCG apparent magnitude estimate of \citetalias{Sand+04} is similar to ours (in a slightly different band), they estimate $R_e=12$ kpc (converting their published value to our adopted cosmology), a factor $\sim 2$ smaller than our estimate. Moreover, their BCG velocity dispersion values, obtained with single slit observations with \texttt{ESI@Keck~II}, are significantly smaller than our measurements, obtained with \texttt{MUSE@VLT} integral field observations \citepalias[compare Fig.~7 and Table~5 in][with Fig.~\ref{f:vdp}]{Sand+04}. \citetalias{Sand+04}'s BCG velocity dispersion values are 54 and, respectively, 119 $\ks$ below our values in the radial ranges where they made their estimates, 0-6 and 6-28 kpc, respectively, corresponding to a 18\% and 32\% under-estimate, respectively. The superior quality of our \texttt{MUSE} measurements appears therefore crucial in achieving an accurate $\gdm$ determination.

Our $\gdm$ value for MACS~1206 is somewhat below the NFW profile value, while we found almost exactly the NFW value for the cluster Abell~S1063 that we analysed with the same methodology in \citet[][$\gdm=0.99 \pm 0.04$]{Sartoris+20}.
Part of the difference among these values could be related to the general assumption of spherical symmetry combined with different levels of triaxiality and different orientations with respect to the line-of-sight. However, the study of \citet{Sereno+18} finds very similar triaxial shapes and orientations for the two clusters, so the difference in their $\gdm$ values is probably intrinsic.

According to \citet{Sereno+18}, the orientation of the major axis of MACS~1206 with respect to the line-of-sight is 78$^{\circ}$, and this is also suggested by the very elliptical projected shape of the BCG. The study of \citet{He+20}, based on the \texttt{C-EAGLE} numerical simulations, shows that the value of $\gdm$ inferred under the spherical symmetry assumption can differ from the intrinsic value by $\pm 0.2$ rms, with a negative bias of $-0.2$ when the BCG is viewed along its minor axis \citep[see Fig.~15 in][]{He+20}. The intrinsic value of $\gdm$ of MACS~1206 could then be even closer to the NFW value than we find in our analysis. In any case our
$\gdm$ value for MACS~1206 is in agreement with the expectations for massive halos from the \texttt{C-EAGLE} simulations of \citet{He+20}, which suggests a
decreasing trend of $\gdm$ with halo mass \citep[halos as massive as MACS~1206 have $\gdm \approx 0.8$ rather than 1, see][]{He+20}.

Unlike the high $\gdm$ values of Abell~S1063 and MACS~J0416$-$2403, our result is not in contrast with the observational results of \citet{Newman+13b}, who found an average value of $\gdm=0.50 \pm 0.13$ over seven clusters (we show these $\gdm$ values in Fig.~\ref{f:gammas}). However, \citet{Newman+13b}'s value is in significant tension with pure-CDM halo inner slope values, while our value for MACS~1206 is not.

The intermediate $\gdm$ value we find for MACS~1206 therefore removes the tension with CDM models that was created by the previous measurement of \citetalias{Sand+04}, and suggested by the analysis of \citet{LBJ22},
but it does not allow to reject alternative DM models to CDM, such as the self-interacting DM model of \citet{SS00}. The difference with the $\gdm$ value of the Abell S1063 cluster, that we analysed with the same technique and data of the same quality, suggests that different physical processes have been or are now at work in the central regions of these two clusters, shaping the inner slope of the DM profile, or that we observe the two clusters at different stages of their evolution. Clearly, precise $\gdm$ determinations for more clusters are needed to investigate this topic.

\section{Summary and conclusions}\label{s:summ}
We determined the total mass profile of the $z=0.44$ massive cluster MACS~1206, from 2 kpc to 2 Mpc, and, separately, its DM and baryonic components. This result was obtained by applying an extension of the \texttt{MAMPOSSt} method \citep{MBB13,Pizzuti+23} to maximize the combined likelihoods of the observed BCG velocity dispersion profile and the velocity distribution of cluster member galaxies. Our total mass profile is in remarkable agreement with independent determinations based on X-ray observations and strong lensing \citet{Caminha+17}, after accounting for the mass contribution of a foreground group of galaxies to the strong lensing signal. This comparison also shows no significant evidence of an hydrostatic bias, often claimed in X-ray cluster mass profile determinations.

As an additional output of our analysis, we constrained the BCG stellar velocity anisotropy to be close to isotropic, as expected based on previous works on giant ellipticals \citep{Kronawitter+00,Santucci+23}. We found that the orbits of cluster galaxies are radially elongated, increasingly with radius, a feature common to most clusters \citep[e.g.][]{NK96,Biviano+13,Mamon+19,Biviano+21}, 

Our main result is the determination of the DM profile inner slope, $\gdm=0.7_{-0.1}^{+0.2}$ (68\% c.l.) $\pm 0.3$ (95\% c.l.), the most accurate determination for this cluster so far. Our value is somewhat below the NFW profile slope predicted in CDM simulations, but fully consistent with it when considering the scatter in the inner slopes of DM halos and the decreasing trend of $\gdm$ with halo mass \citep{He+20}. Our value does not support the claim of a cored inner mass distribution for this cluster by \citet{LBJ22}, and it
is significantly larger than the $\gdm$ measurement by \citetalias{Sand+04} for the same cluster, a difference that we attribute to the significantly increased accuracy and precision of our BCG velocity dispersion profiles measurement. 

Our results for the massive clusters MACS~1206 and Abell~S1063 \citep[previously analysed with the same methodology by] []{Sartoris+20} appear in very good agreement with state-of-the-art $\Lambda$CDM numerical simulations, at variance with the findings of \citetalias{Sand+04} and \citet{Newman+13b}. More high-quality measurement of $\gdm$ are needed to determine its distribution across different clusters, and, in combination with cluster properties, understand the physical origin of its scatter.

\begin{acknowledgments}
We dedicate this work to the memory of our beloved friend and esteemed colleague Mario Nonino.
We thank Carlos Frenk for useful discussion.  We thank the anonymous referee for her/his report that helps improving this paper.
We acknowledge financial support through grants PRIN-MIUR 2017WSCC32. AB acknowledges the financial contribution from the INAF mini-grant 1.05.12.04.01 {\it "The dynamics of clusters of galaxies from the projected phase-space distribution of cluster galaxies"}.
LP acknowledges support from the Czech Academy of Sciences under the grant number LQ100102101. SE acknowledges the financial contribution from the contracts ASI-INAF Athena 2019-27-HH.0, {\it ``Attivit\`a di Studio per la comunit\`a scientifica di Astrofisica delle Alte Energie e Fisica Astroparticellare''} (Accordo Attuativo ASI-INAF n. 2017-14-H.0), and from the European Union’s Horizon 2020 Programme under the AHEAD2020 project (grant agreement n. 871158).
\end{acknowledgments}

\bibliography{master}{}
\bibliographystyle{aasjournal}
\end{document}